\documentclass[twocolumn]{aastex631}

\usepackage{amsmath}
\begin{document}

\title{Testing $f(R)$ Gravity from Cosmic Shear Measurements}

\author{Jiachen Bai}\thanks{astrobaijc@gmail.com}
\affiliation{School of Physics and Astronomy, Beijing Normal University, Beijing 100875, P.R.China}

\author{Jun-Qing Xia}\thanks{xiajq@bnu.edu.cn}
\affiliation{School of Physics and Astronomy, Beijing Normal University, Beijing 100875, P.R.China}
\affiliation{Institute for Frontiers in Astronomy and Astrophysics, Beijing Normal University, Beijing, 102206, P.R.China}

\author{Gong-Bo Zhao}
\affiliation{National Astronomical Observatories, Chinese Academy of Sciences, Beijing, 100101, P.R.China}
\affiliation{School of Astronomy and Space Science, University of Chinese Academy of Sciences, Beijing, 100049, P.R.China}
\affiliation{Institute for Frontiers in Astronomy and Astrophysics, Beijing Normal University, Beijing, 102206, P.R.China}



\begin{abstract}
In this work, we perform a detailed analysis to constrain the Hu-Sawicki \(f(R)\) gravity model, using cosmic shear data from three prominent Stage-III weak lensing surveys: DES-Y3, KiDS-1000, and HSC-Y3. To accurately model the nonlinear matter clustering in the analysis of cosmic shear signals, we employ \texttt{FREmu}, a recently developed power spectrum emulator for the \(f(R)\) gravity trained on the Quijote-MG simulations. This emulator achieves precise predictions, limiting the errors to 5\% on scales of \(0.009h\,{\rm Mpc}^{-1} < k < 0.5h\,{\rm Mpc}^{-1}\). Our findings reveal that cosmic shear data alone impose only weak constraints on the \(f(R)\) parameter \(\log_{10}|f_{R_0}|\). To improve these constraints, we incorporate state-of-the-art external observations, including data from the cosmic microwave background and baryon acoustic oscillations. The inclusion of these external datasets significantly enhances the constraints,  yielding an upper limit of \(\log_{10}|f_{R_0}| < -4.98\) at the 95\% confidence level.

\end{abstract}

\keywords{Modified Gravity --- Large-Scale Structure --- Weak Lensing --- Cosmology }

\section{Introduction}



Testing the fundamental nature of gravity remains a cornerstone objective in contemporary physics research. As the principal force governing the growth and evolution of large-scale structures (LSS) in the Universe, gravity can be directly investigated through detailed cosmological observations. The advent of increasingly precise datasets and innovative experimental methodologies has propelled modern cosmology into a new era characterized by unprecedented theoretical and modeling sophistication \citep{2022ARNPS..72....1T}. These advancements render the testing of gravity not only viable but also a highly promising pathway for deepening our understanding of fundamental physics.


From a theoretical perspective, significant advancements in computational techniques have considerably enhanced the precision and efficiency of predictions for LSS observables. Multi-sample simulations across various cosmological models now enable highly accurate predictions of LSS properties \citep{2020ApJS..250....2V}. Additionally, the integration of machine learning methods has accelerated the prediction process, facilitating a more seamless connection between theoretical frameworks and observational data.


From an observational standpoint, Stage-III LSS surveys, such as the Dark Energy Survey (DES; \citealt{2016MNRAS.460.1270D}), the Kilo-Degree Survey (KiDS; \citealt{2013ExA....35...25D}), and the Hyper Suprime-Cam Survey (HSC; \citealt{2018PASJ...70S...4A}), have yielded high-quality datasets. These surveys focus on cosmic shear, a weak gravitational lensing (WL) effect that probes the underlying matter distribution. The utilization of LSS observables allows for effective testing of the nature of gravity and the validity of cosmological models \citep{DES:2021zdr, DES:2022ccp, Joudaki:2016kym, SpurioMancini:2019rxy, 2021PhRvD.104h3527V}.


Over the past two decades, the \(\Lambda\)CDM model has emerged as the most rigorously tested cosmological framework. However, alternative cosmological models, including modified gravity (MG) theories, still require robust testing and constraints. This work focuses on the Hu-Sawicki \(f(R)\) gravity model \citep{2007PhRvD..76f4004H}, which offers an alternative explanation for cosmic acceleration without invoking a cosmological constant \(\Lambda\). This model predicts distinct deviations in LSS observables, particularly at smaller scales (\(k > 0.1 h\,\mathrm{Mpc}^{-1}\)).


Linear-scale analytical tools, such as \texttt{MGCAMB} \citep{2023JCAP...08..038W} and \texttt{MGCLASS} \citep{2022JCAP...05..030S}, are insufficient for analyzing nonlinear scales required by real-world observations. While simulations can address these scales, they are computationally expensive. To overcome this limitation, machine learning-based emulators, such as \texttt{FREmu} \citep{2024ApJ...971...11B}, provide efficient predictions for the nonlinear matter power spectrum in \(f(R)\) gravity.



Our analysis demonstrates that using cosmic shear data from Stage-III surveys alone imposes limited constraints on \(f(R)\) gravity due to the model's sensitivity and the current precision of observational datasets. To improve these constraints, we incorporate external datasets, including the cosmic microwave background (CMB) and baryon acoustic oscillations (BAO). While \(f(R)\) gravity remains unconstrained with external datasets alone, their combination with LSS data significantly enhances the constraints on the \(f(R)\) parameter, \(\log_{10}|f_{R_0}|\). These tests highlight the potential of joint-probe analyses in constraining modified gravity theories.


This article is organized as follows: Section 2 provides a comprehensive overview of the theoretical foundations of this study, including \(f(R)\) gravity, emulation techniques, cosmic shear, and associated systematics. Section 3 introduces the observational datasets used, including cosmic shear data from individual surveys and complementary external datasets. The methods employed in the analysis are also detailed in this section. In Section 4, we present the analysis process, along with the results of the constraints on \(f(R)\) gravity.  Finally, we summarize the findings and discuss potential future directions for research in Section 5. 

\section{Theory}

\subsection{The Hu-Sawicki $f(R)$ Gravity}

The principle of least action governs gravitational dynamics in $f(R)$ gravity through a modification of the Einstein-Hilbert action, introducing an additional $f(R)$ term:
\begin{equation}
S=\int \mathrm{d}^4 x \sqrt{-g}\left[\frac{R+f(R)}{16 \pi G}+\mathcal{L}_{\mathrm{m}}\right],
\end{equation}
where $g$ is the determinant of the metric tensor $g_{\mu\nu}$, $R$ is the Ricci scalar, $G$ is the gravitational constant, and $\mathcal{L}_{\mathrm{m}}$ denotes the matter Lagrangian density.

Varying the action with respect to $g_{\mu\nu}$ yields the modified field equations:
\begin{equation}
G_{\mu\nu} + f_{R}R_{\mu\nu} - \left(\frac{f}{2} - \Box f_{R}\right) g_{\mu\nu} - \nabla_{\mu} \nabla_{\nu} f_{R} = \kappa^{2} T_{\mu\nu},
\end{equation}
where $f_R = \frac{\mathrm{d}f}{\mathrm{d}R}$ introduces a scalar degree of freedom (SDOF).

Under the quasi-static approximation, these field equations reduce to the modified Poisson equation and the equation of motion for the SDOF:
\begin{align}
    &\boldsymbol{\nabla}^2 \Phi = \frac{16 \pi G}{3} \delta \rho - \frac{1}{6} \delta R, \\
    &\boldsymbol{\nabla}^2 f_R = \frac{1}{3}(\delta R - 8 \pi G \delta \rho),
\end{align}
where $\Phi$ is the Newtonian potential, and $\delta \rho$ and $\delta R$ are perturbations in matter density and the Ricci scalar, respectively.

The Hu-Sawicki model parameterizes $f(R)$ as:
\begin{equation}
    f(R) = -m^2 \frac{c_1 \left(R / m^2\right)^n}{c_2 \left(R / m^2\right)^n + 1},
\end{equation}
where $m = H_0 \sqrt{\Omega_{\rm m}}$, and $c_1$, $c_2$, and $n$ are free parameters. This parameterization for \(f(R)\) gravity exhibits an asymptotic behavior in the high-curvature regime, effectively mimicking a cosmological constant in the \(\Lambda\mathrm{CDM}\) model. To reproduce the \(\Lambda\mathrm{CDM}\) background expansion history, the free parameters \(c_1\) and \(c_2\) must satisfy the following relation:
\begin{equation}
    \frac{c_1}{c_2} = 6\, \frac{\Omega_{\Lambda}}{\Omega_{\mathrm{m}}}
\end{equation}

In the high-curvature limit, the functional form of \(f(R)\) can be approximated as:
\begin{equation}
    f(R) \approx -2\Lambda + \frac{f_{R_0}}{n}\, \frac{R_0^{n+1}}{R^n}.
\end{equation}

In this work, we fix the parameter \(n = 1\). As \(|f_{R_0}| \rightarrow 0\), the model asymptotically approaches \(\Lambda\mathrm{CDM}\), while even small deviations from General Relativity can significantly impact the growth of large-scale structure.

\subsection{The Emulation of Nonlinear Power Spectra}
The theoretical prediction of nonlinear matter power spectra for \(f(R)\) gravity is a crucial component of this study, as it underpins the computation of the cosmic shear data vector. While the linear matter power spectra for \(f(R)\) gravity can be computed analytically using Einstein-Boltzmann solvers such as \texttt{MGCAMB} \citep{2023JCAP...08..038W} or \texttt{MGCLASS} \citep{2022JCAP...05..030S}, capturing nonlinear effects requires resource-intensive N-body simulations. However, the computational demands of such simulations make them impractical for general use and inefficient for parameter inference, which involves sampling across numerous parameter combinations.  

To address this challenge, we employ a simulation-based emulator, \texttt{FREmu} \citep{2024ApJ...971...11B}, to predict the nonlinear matter power spectra for \(f(R)\) gravity. \texttt{FREmu} leverages machine learning algorithms, including Artificial Neural Networks (ANN) and Principal Component Analysis (PCA), to accelerate predictions. It builds upon the approximations provided by \texttt{HALOFIT} \citep{2021MNRAS.502.1401M}, integrated within \texttt{CAMB} \citep{2011ascl.soft02026L}, and is trained using the Quijote-MG simulation suite \citep{baldi2024}.  

The Quijote-MG simulations cover a wide range of cosmological parameters, including the matter density parameter \(\Omega_{\mathrm{m}}\), baryon density parameter \(\Omega_{\mathrm{b}}\), Hubble parameter \(h\), spectral index \(n_s\), primordial power spectrum amplitude \(A_\mathrm{s}\), the sum of neutrino masses \(M_\nu\), and the \(f(R)\) parameter \(f_{R_0}\). Importantly, we adopt \(A_\mathrm{s}\) as a cosmological input parameter because \(\sigma_8\), the root-mean-square fluctuation of the power spectrum, is influenced by \(f_{R_0}\) in \(f(R)\) gravity and cannot be independently determined but must instead be derived. Using \(A_\mathrm{s}\) provides greater clarity and consistency in the parameterization. \textbf{The simulations were performed with the MG-GADGET code \citep{Puchwein:2013lza}, which evolves the N-body dynamics in \(f(R)\) gravity and self-consistently accounts for the chameleon screening mechanism.}

To combine predictions across scales, we rely on \texttt{MGCAMB} for large-scale power spectra (\(k < 0.09 \, h/\mathrm{Mpc}\)) and \texttt{FREmu} for smaller scales (\(k \geq 0.09 \, h/\mathrm{Mpc}\)). This hybrid approach ensures accurate modeling across the full range of relevant scales. It is also noteworthy that \texttt{FREmu} is credible up to a maximal wavenumber of \(k \approx 0.5\, h/\mathrm{Mpc}\), beyond which its predictions become unreliable. Accordingly, scale cuts are imposed in the following analysis, as detailed in Sec.~\ref{shear}.

The Quijote-MG suite includes only dark matter and massive neutrino particles, while the actual Universe also contains baryonic matter, which significantly impacts the matter power spectrum, particularly at small scales. To incorporate baryonic effects, we adopt the Baryonic Correction Model (BCM) proposed by \citet{2015JCAP...12..049S}, which introduces two additional parameters, \(M_\mathrm{c}\) and \(\eta_\mathrm{BCM}\). While the BCM provides an effective means of accounting for baryonic effects, we apply appropriate scale cuts (detailed in Sec.~\ref{data}) to mitigate inaccuracies that arise at smaller scales where the BCM may fail.

\begin{figure}
    \centering
    \includegraphics[width=\linewidth]{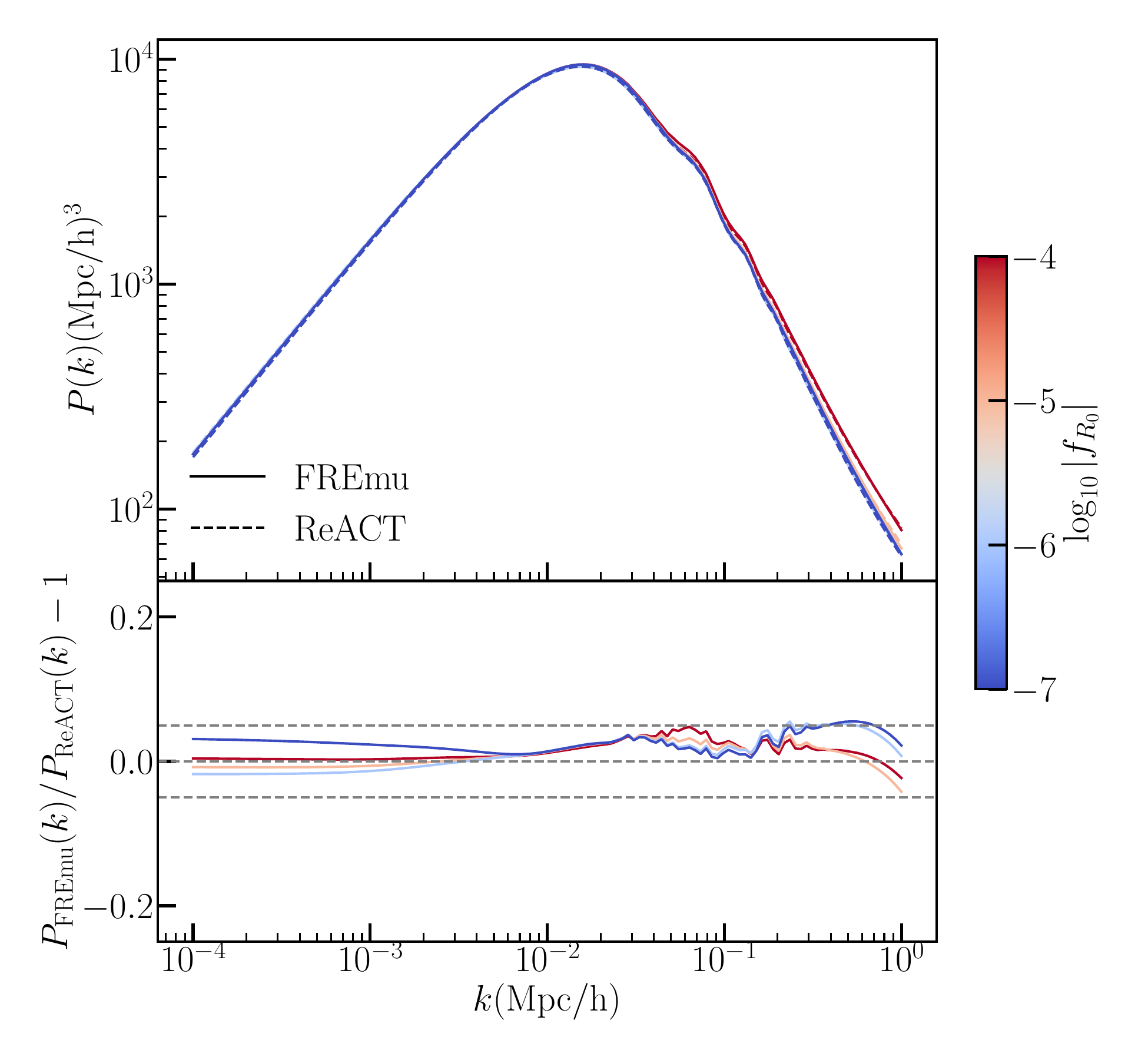}
    \caption{Validation of the emulator performance by comparison with \texttt{ReACT} at redshift \(z=1\). Four samples with \(\log_{10}|f_{R_0}| = -4, -5, -6, -7\) are selected. The relative errors remain within 5\%.}
    \label{fig:validation}
\end{figure}

To validate the performance of the emulator, we compare it against another power spectrum emulator, \texttt{ReACT}~\citep{Cataneo:2018cic, Bose:2020wch, Bose:2021mkz}. The comparison is performed at \(z=1\), which approximately corresponds to the mean redshift of cosmic shear observations. In Fig.~\ref{fig:validation}, we select four representative samples with \(\log_{10}|f_{R_0}| = -4, -5, -6, -7\). The relative errors remain within 5\%, indicating that the predictions from our emulator are reliable.

\subsection{Cosmic Shear}

The angular power spectrum of galaxy ellipticity correlations caused by WL is directly related to an integral over the nonlinear matter power spectrum weighted by the lensing kernel following the flat-sky and Limber approximations \citep{1953ApJ...117..134L} as:
\begin{equation}
C_{\kappa \kappa}^{i j}(\ell)=\int_0^{\chi_\mathrm{H}} {\rm d} \chi \frac{q_\kappa^i(\chi) q_\kappa^j(\chi)}{\chi^2} P_\mathrm{m}\left[\frac{\ell+1 / 2}{\chi}, z(\chi)\right],
\end{equation}
where the Latin indices $(i/j)$ label the tomographic redshift bins, $\ell$ is the angular multipole, $\chi$ denotes the comoving radial distance, $\chi_\mathrm{H}$ represents the comoving horizon distance, while $z$ and $k$ are redshift and modulus of the wave vector respectively. The lensing kernel $q_\kappa^i(\chi)$ can be expressed as:
\begin{equation}
q^i_\kappa(\chi)=\frac{3 H_0^2 \Omega_m}{2 c^2} \frac{\chi}{a(\chi)} \int_\chi^{\chi_\mathrm{H}} \mathrm{d} \chi^{\prime} \frac{\mathrm{d} z}{\mathrm{~d} \chi^{\prime}} \frac{n^i\left[z\left(\chi^{\prime}\right)\right]}{\bar{n}^i} \frac{\chi^{\prime}-\chi}{\chi^{\prime}}.
\end{equation}
In the above, $n^i(z)$ denotes the redshift distribution of galaxies in the $i$-th tomographic bin, $a$ is the scale factor, $\bar{n}^i$ represents the mean number density of source galaxies. $H_0$ is the Hubble constant, $\Omega_m$ is the matter fraction of the Universe and $c$ is the speed of light.

The two-point correlation functions (2PCFs) for two redshift bins can be expressed in terms of the convergence power spectrum $C_{\kappa \kappa}^{i j}(\ell)$ as:
\begin{equation}
\xi_{ \pm}^{i j} (\theta)=\sum_{\ell} \frac{2 \ell+1}{2 \pi \ell^2(\ell+1)^2}\left[G_{\ell, 2}^{+}(\cos \theta) \pm G_{\ell, 2}^{-}(\cos \theta)\right] C_{\kappa \kappa}^{i j}(\ell)
\end{equation}
where $G_{\ell, 2}^{\pm}(\cos \theta)$ are computed from Legendre polynomials.

\subsection{Systematics Modeling}
To model intrinsic alignment (IA) effects caused by the intrinsic shapes of galaxies, we follow the non-linear alignment model (NLA; \citealt{2004PhRvD..70f3526H}, \citealt{2007NJPh....9..444B}). The IA effects are accounted for by incorporating contributions from intrinsic-intrinsic (II) and gravitational-intrinsic (GI/IG) terms, which modify the observed cosmic shear signal. These effects are parameterized using three additional parameters: the IA amplitude $A_{\mathrm{IA}}$, which quantifies the overall strength of intrinsic alignment; the redshift scaling parameter $\eta_{\mathrm{IA}}$, describing how the IA amplitude evolves with redshift and the pivot redshift, $z_0$, specific to each survey, serving as the reference point for the redshift evolution of IA effects.

To address uncertainties in the redshift distributions of galaxies, we introduce nuisance parameters, $\Delta z^i$, which shift the redshift distribution of each bin by a specified amount. These shifts reflect potential calibration biases in the photometric redshift estimates, ensuring a more robust characterization of the galaxy distribution. Furthermore, we account for potential biases in shear measurements by including multiplicative bias parameters, $m^i$, for each redshift bin. These parameters correct for systematic errors in shear calibration, enabling a more accurate extraction of the true cosmic shear signal. By incorporating these systematic effects, we enhance the reliability of our analysis and ensure that our constraints on cosmological and modified gravity parameters are not skewed by observational biases.

\section{Data and Methods}\label{data}

\subsection{Cosmic Shear from Weak Lensing Surveys}\label{shear}

In this study, we focus on cosmic shear data from three significant weak lensing (WL) surveys: the Dark Energy Survey Year 3 (DES-Y3; \citealt{2022PhRvD.105b3514A}), the fourth data release of the Kilo-Degree Survey (KiDS-1000; \citealt{2021A&A...645A.104A}), and the Hyper Suprime-Cam Year 3 (HSC-Y3; \citealt{2023PhRvD.108l3518L}). These datasets form the foundation of our analysis, which aims to constrain the \(f(R)\) gravity model.

{\bf DES-Y3:} The Dark Energy Survey Year 3 data release (DES-Y3) encompasses approximately three years of observational data, covering around 5,000 square degrees of the southern sky. The DES-Y3 data includes deep, multi-band photometry, which enables precise measurements of galaxy shapes, clustering, and weak gravitational lensing signals. In this work, we only utilize its cosmic shear measurements for analysis.

{\bf DES-Y3 Blue:} Additionally, we incorporate the recent DES-Y3 Blue dataset \citep{2024arXiv241022272M}, which focuses exclusively on blue, star-forming galaxies. Since the IA effects are negligible for this population, we can analyze the dataset without applying scale cuts related to IA effects, thereby maximizing the statistical power of the cosmic shear data.

{\bf KiDS-1000:} The Kilo-Degree Survey's KiDS-1000 dataset includes high-quality imaging data covering roughly 1,000 square degrees. Designed for weak gravitational lensing studies, KiDS-1000 facilitates probing the distribution of dark matter and cosmic structure growth, making it a valuable dataset for cosmological analyses.

{\bf HSC-Y3:} The Hyper Suprime-Cam Subaru Strategic Program Year 3 (HSC-Y3) dataset represents the latest release from the HSC-SSP survey. With approximately 1,000 square degrees of coverage, it provides some of the deepest and highest-quality imaging data among current WL surveys, offering critical insights into weak lensing phenomena.

{\bf DES-Y3 + KiDS-1000:} The footprints of these three WL surveys partially overlap, introducing cross-covariance between data vectors from any two surveys, even when their measurements are independent. To avoid introducing cross-covariance between datasets with overlapping footprints, we combine the DES-Y3 and KiDS-1000 datasets by excluding the overlapping spatial regions in DES-Y3 \citep{2023OJAp....6E..36D}. This ensures that the combined dataset is independent and free from cross-covariance effects. Notably, we exclude HSC-Y3 from the hybrid analysis due to its overlapping regions with DES-Y3 and KiDS-1000. Handling such overlaps by excluding regions from the skymap is beyond the scope of this work. Therefore, the hybrid dataset only includes DES-Y3 and KiDS-1000 data.

While all three surveys provide cosmic shear data vectors spanning scales from the largest to the smallest scales, fiducial analyses apply scale cuts to mitigate uncertainties from small-scale effects, baryonic physics, and modeling inaccuracies. To ensure the robustness of our results, we first adopt the fiducial scale cuts defined in the official analyses of the respective surveys. We then further exclude those data points that do not satisfy the following criterion: \begin{equation} 
    \left(\xi_{ \pm, \mathrm{full} }^{ij} - \xi_{ \pm, k<0.5, \mathrm{Mpc}/h }^{ij}\right) < \eta_{\mathrm{th}}\times \sigma_{ \pm }^{ij} .
\end{equation} 

Here, $\xi_{ \pm, \mathrm{full} }^{ij}$ denotes the predicted cosmic shear signal from the emulator using the full matter power spectrum, while $\xi_{ \pm, k<0.5, \mathrm{Mpc}/h }^{ij}$ is the corresponding prediction excluding contributions from scales smaller than $k=0.5, \mathrm{Mpc}/h$. The quantity $\sigma_{ \pm }^{ij}$ represents the standard deviation of the data point, computed from the diagonal elements of the covariance matrices provided by each survey. The parameter $\eta_{\mathrm{th}}$ defines the tolerance threshold, which we set to 5\%.

Notably, both the KiDS and HSC collaborations have previously attempted to constrain $f(R)$ gravity \citep{2021A&A...649A..88T,2021PhRvD.104h3527V}, but did not provide definitive constraints on $\log_{10}|f_{R_0}|$ using weak lensing data alone. Similarly, we find that shear-only datasets yield comparable results, emphasizing the importance of incorporating external datasets to improve constraints on $\log_{10}|f_{R_0}|$. This study demonstrates the utility of hybrid datasets in addressing limitations in individual survey analyses and paves the way for more precise constraints on modified gravity theories.


\subsection{External Datasets}

To enhance the constraining power on \(f(R)\) gravity, we incorporate additional external datasets that complement the weak lensing measurements and provide more stringent constraints on cosmological parameters.

We utilize data from the Planck satellite’s 2018 data release, including the low-multipole (low-$\ell$) temperature anisotropy and E-mode (EE) polarization likelihoods \citep{2020A&A...641A...5P}. To improve sensitivity to small-scale CMB fluctuations, we incorporate the latest Planck PR4 data products \citep{2022MNRAS.517.4620R}, specifically the CamSpec high-multipole (high-$\ell$) TTTEEE likelihoods and the most recent CMB lensing data from \citet{2022JCAP...09..039C}.

Additionally, we include the latest Baryon Acoustic Oscillation (BAO) measurements from the Dark Energy Spectroscopic Instrument (DESI) \citep{2024arXiv240403002D, DESI:2025zgx}. These measurements enhance the precision of distance constraints at various redshifts, offering complementary insights into the large-scale structure of the Universe and the cosmic expansion history.

To conclude this section, the external datasets include PR4 CMB and DESI BAO. In the subsequent sections, we refer to the combination of these datasets as "All Ext."

\begin{table}
\centering
\caption{Priors used in the fiducial cosmic shear analysis. The table lists the priors for cosmological parameters, baryonic feedback parameters, and intrinsic alignment parameters. Uniform priors $U(a, b)$ indicate flat distributions within the specified ranges. The systematic parameters are handled with survey-specific priors, following the standards provided by each dataset.}
\label{tab:prior}
\begin{tabular}{|l l|}
\hline
\multicolumn{2}{|c|}{\textbf{Cosmology}} \\ \hline
$\Omega_\mathrm{m}$ & $U(0.1, 0.5)$ \\ 
$\Omega_\mathrm{b}$ & $U(0.03, 0.07)$  \\ 
$H_0$ & $U(50, 90)$  \\ 
$n_\mathrm{s}$ & $U(0.8, 1.2)$  \\ 
$\log(10^{10}A_\mathrm{s})$ & $U(1.61, 3.91)$  \\ 
$\sum m_\nu$ [eV] & 0.06 (fixed) \\ 
$\log_{10}|f_{R_0}|$  & $U(-7, -3.5)$ \\ \hline \hline
\multicolumn{2}{|c|}{\textbf{Baryonic Feedback}} \\ \hline
$\log_{10}M_\mathrm{c}$ & $U(12, 16)$ \\
$\eta_\mathrm{BCM}$ & $U(0.1, 1)$ \\ \hline \hline
\multicolumn{2}{|c|}{\textbf{Intrinsic Alignment}} \\ \hline
$A_\mathrm{IA}$ & $U(-5, 5)$ \\
$\eta_\mathrm{IA}$ & $U(-5, 5)$ \\ 
$z_0$ & 0.62 (0.3 for KiDS-1000) \\ \hline
\end{tabular}
\end{table}

\begin{figure*}[t]
    \centering
    \includegraphics[width=1.8\columnwidth]{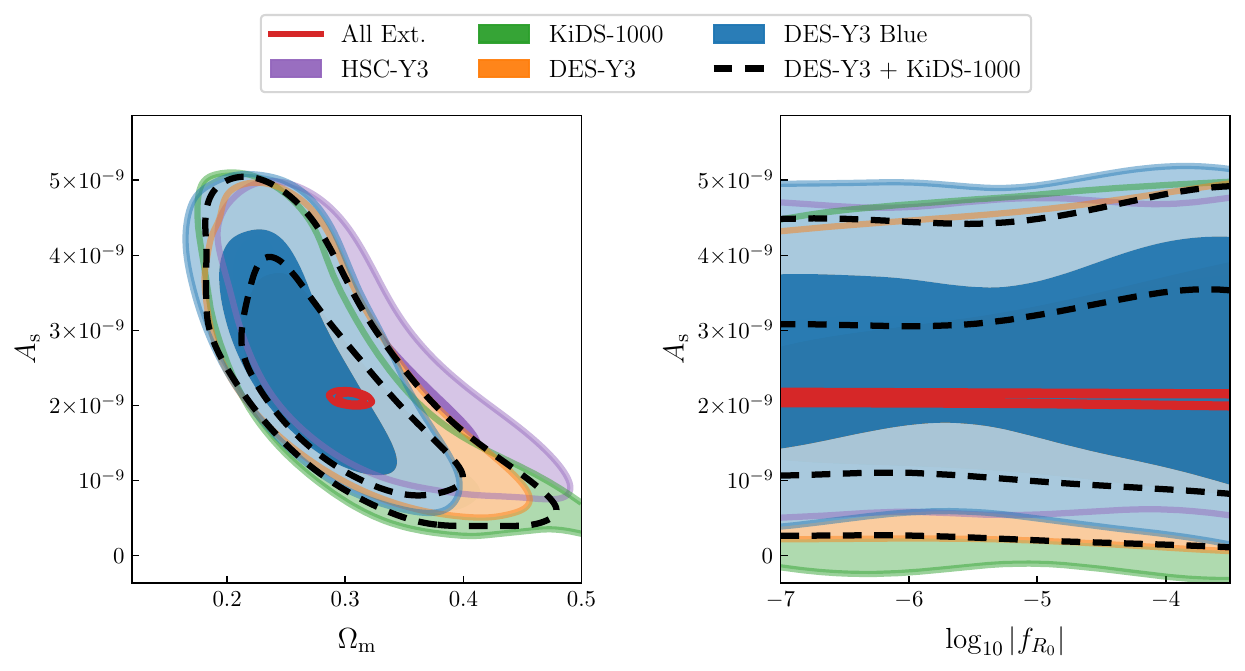}
    \caption{Comparison of cosmic shear constraints on $A_\mathrm{s}$ versus $\Omega_\mathrm{m}$ (left) and $A_\mathrm{s}$ versus $\log_{10}|f_{R_0}|$ (right) from five datasets: DES-Y3, DES-Y3 Blue, KiDS-1000, HSC-Y3, and a combined DES-Y3 + KiDS-1000 analysis, with constraints derived from external datasets shown in red.}
    \label{fig:shear_only}
\end{figure*}

\subsection{Method: MCMC and Implementation}

In this work, we employ the Markov Chain Monte Carlo (MCMC) method for parameter estimation \citep{1970Bimka..57...97H}, which is a well-established statistical technique for sampling from the posterior distribution of cosmological parameters. We compute the posterior distribution using Bayes' theorem, where the likelihood function plays a central role in determining the probability of observing the data given a specific set of model parameters. The likelihood function here is given by:
\begin{equation}
\mathcal{L}(\theta |D)\propto\exp \left[-\frac{1}{2}(\mathbf{O}(D)-\mathbf{M}(\theta))^T \mathbf{\Sigma}^{-1}(\mathbf{O}(D)-\mathbf{M}(\theta))\right].
\end{equation}

Here, $\theta$ denotes the model parameters, $D$ represents the observed data, $\mathbf{O}(D)$ represents a column vector of observables derived from the data, $\mathbf{M}(\theta)$ represents a column vector of model predictions $M_i(\theta)$, and $\Sigma$ denotes the covariance matrix of the measurement errors.

When combining multiple datasets in our analysis (such as cosmic shear and CMB data), we assume the datasets are independent. Thus, the total log-likelihood is simply the sum of the individual log-likelihoods from each dataset:
\begin{equation}
\log \mathcal{L}(\theta | D_\mathrm{Joint})=\log \mathcal{L}(\theta | D_\mathrm{WL}) + \log \mathcal{L}(\theta | D_\mathrm{All\,Ext}).
\end{equation}

For the MCMC sampling, we use the algorithm proposed by \citet{2013PhRvD..87j3529L}, implemented in the \texttt{Cobaya} package \citep{2021JCAP...05..057T}. This package provides an efficient framework for cosmological parameter estimation and is compatible with the latest Planck PR4 datasets.

Additionally, we write our own version of the likelihood code for the three latest WL datasets in \texttt{Cobaya}. The data vectors of 2PCFs are computed with \texttt{PyCCL} \citep{2019ApJS..242....2C}. For the $f(R)$ gravity model, the nonlinear power spectra are provided by the \texttt{FREmu} emulator, which is integrated into \texttt{PyCCL}. In the CMB pipeline, the linear power spectra are computed with \texttt{MGCAMB}.

\begin{figure*}[t]
    \centering
	\includegraphics[width=1.8\columnwidth]{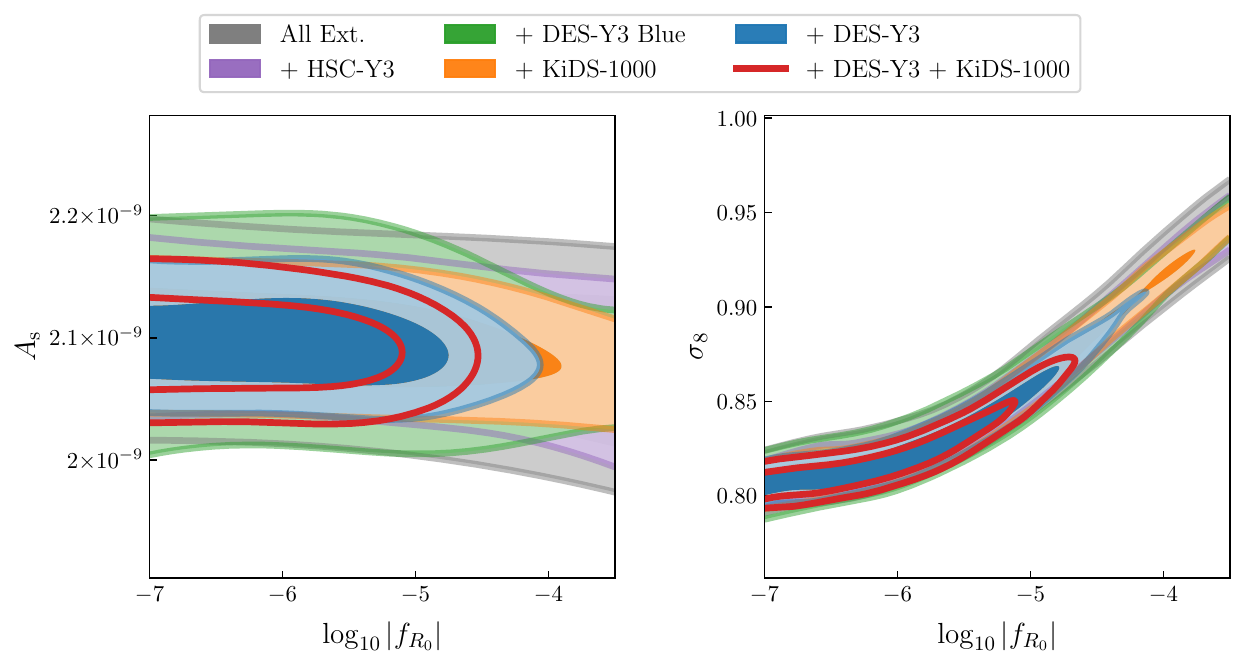}
    \caption{Constraints on $\log_{10}|f_{R_0}|$ from the combination of cosmic shear and external datasets. The left panel depicts $\log_{10}|f_{R_0}|$ versus $A_\mathrm{s}$ and the right panel shows the relationship between $\log_{10}|f_{R_0}|$ and $\sigma_8$, where the inclusion of cosmic shear enhances the precision of the constraints, with the tightest limit obtained from the joint analysis of DES-Y3 + KiDS-1000 + All Ext.}
    \label{fig:shear_cmb}
\end{figure*}

\section{Analyses and results}
\subsection{Cosmic Shear Only}
In this study, we examine \(f(R)\) gravity by conducting an analysis of cosmic shear data obtained from five principal datasets: DES-Y3 (and the blue shear), KiDS-1000, HSC-Y3, and a combined DES-Y3 + KiDS-1000 dataset. The primary objective of this analysis is to investigate the constraints on the modified gravity parameter \(\log_{10}|f_{R_0}|\) within the framework of LSS observations. 

As outlined in Tab.~\ref{tab:prior}, we impose a comprehensive set of cosmological and astrophysical priors on various parameters, encompassing seven cosmological parameters, two baryonic feedback parameters, and three intrinsic alignment parameters. While the neutrino mass parameter is permitted to vary within the \texttt{FREmu} framework, it is held fixed at \(0.06 \, \mathrm{eV}\) in this analysis to mitigate the degeneracy between the neutrino mass and \(\log_{10}|f_{R_0}|\). This degeneracy, previously discussed by \citet{2014MNRAS.440...75B}, lies beyond the scope of the present study due to limitations in the quality of observational data. 

Additionally, priors associated with parameters addressing systematic uncertainties are specific to each survey. For these parameters, we adopt the fiducial priors as defined by the respective methodologies employed in each survey's official analysis.

We first validated our cosmic shear likelihood pipeline using DES-like mock data generated with fixed input parameters, employing the same covariance matrix as DES-Y3. The results closely aligned with the input values, demonstrating the pipeline's accuracy and reliability. For example, the input value of \(\Omega_\mathrm{m}\) was set to 0.25, with a recovered value of \(0.26^{+0.083}_{-0.078}\) at the 95\% confidence level. Similarly, the scalar amplitude \(10^9A_\mathrm{s}\) was initialized at 3.00, yielding a recovered value of \(3.1^{+1.8}_{-1.5}\). The modified gravity (MG) parameter \(\log_{10}|f_{R_0}|\), set at \(-5.00\), remained unconstrained, as expected given the limitations of shear-only data. These findings confirm the robustness of our cosmic shear pipeline and the reliability of its outputs.

Subsequently, we applied the pipeline to real cosmic shear data from three surveys. The left panel of Fig.~\ref{fig:shear_only} illustrates the resulting constraints, which exhibit similar degeneracy patterns between \(A_\mathrm{s}\) and \(\Omega_\mathrm{m}\) within the \(f(R)\) framework. Notably, the results derived from the blue shear dataset display slightly tighter constraints on these parameters, attributed to the inclusion of small-scale information and the reduction of
the parameter space. Additionally, we present results obtained by combining all external datasets, which yield significantly tighter constraints on \(A_\mathrm{s}\) and \(\Omega_\mathrm{m}\).

The right panel of Fig.~\ref{fig:shear_only} focuses on the constraints for \(\log_{10}|f_{R_0}|\), demonstrating that this parameter remains largely unconstrained regardless of whether cosmic shear-only data or external-only data are employed. These findings highlight the current limitations of cosmic shear measurements from Stage-III surveys in constraining theories of gravity.

\subsection{Joint Results}
To achieve more stringent and comprehensive constraints on the modified gravity parameter \(\log_{10}|f_{R_0}|\), it is imperative to combine cosmic shear measurements with state-of-the-art external datasets. These datasets provide complementary insights into the growth of cosmic structure and the Universe's expansion history, enabling a more holistic analysis.

As depicted in Fig.~\ref{fig:shear_only} and Fig.~\ref{fig:shear_cmb}, external datasets alone constrain most cosmological parameters effectively, owing to their sensitivity to large-scale structure and linear perturbations. However, the parameter \(\log_{10}|f_{R_0}|\) remains unconstrained under these conditions.

The integration of cosmic shear data with external datasets markedly improves the results. As illustrated in Fig.~\ref{fig:shear_cmb}, this combined analysis tightens the constraints on \(\log_{10}|f_{R_0}|\) by harnessing the nonlinear growth sensitivity of cosmic shear alongside the large-scale sensitivity of external datasets. This synergy yields a robust upper limit:  
\begin{equation}
\log_{10}|f_{R_0}| < -4.98 \quad \text{(95\% C.L.)}.
\end{equation}

In the right panel of Fig.~\ref{fig:shear_cmb}, we present the relationship between \(\sigma_8\) and \(\log_{10}|f_{R_0}|\) derived from our analysis. When \(A_\mathrm{s}\) is relatively fixed, \(\sigma_8\) increases with higher values of \(\log_{10}|f_{R_0}|\), reflecting the impact of \(f(R)\) gravity in enhancing matter clustering. This trend underscores the influence of \(f(R)\) gravity, as the weak upper limit on \(\log_{10}|f_{R_0}|\) allows for larger \(\sigma_8\) values within this framework.

To assess the significance of nonlinear signals in cosmic shear data, we apply a linear scale cut to the combined DES-Y3 and KiDS-1000 datasets, removing contributions where predictions differ between linear and nonlinear theories. Our findings indicate that, upon excluding all nonlinear components, the remaining data provide minimal constraining power, even when supplemented with external datasets. This highlights the critical importance of incorporating nonlinear signals to extract meaningful constraints from cosmic shear measurements.

\begin{figure}[t]
    \centering
	\includegraphics[width=1.0\columnwidth]{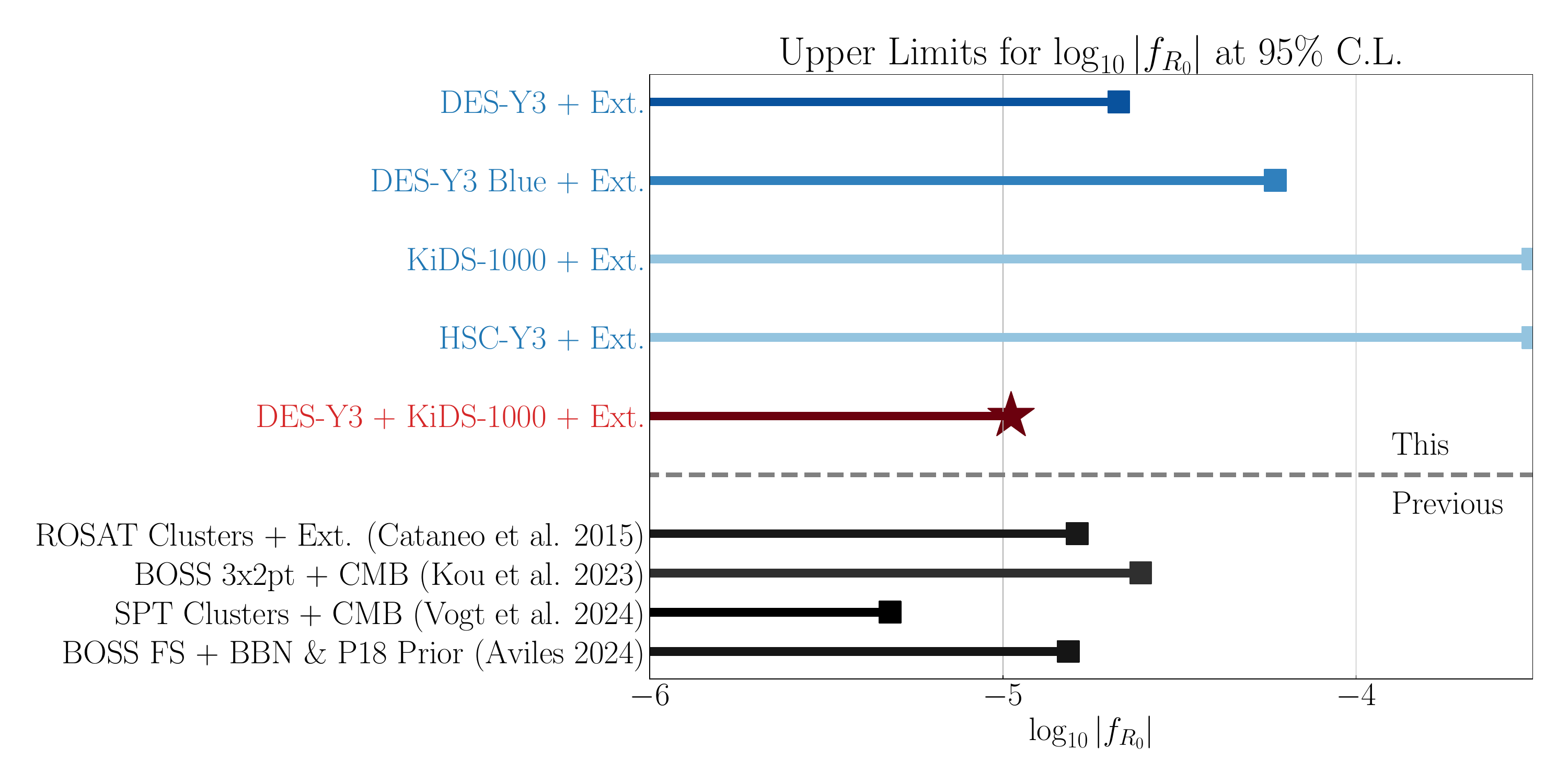}
    \caption{The constraints on $\log_{10}|f_{R_0}|$ from this work are compared with those from previous works. The best result from this work is shown in red, while other constraints using cosmic shear are presented in blue. Results from previous studies are depicted in black.}
    \label{fig:logfR0_bar}
\end{figure}

To provide context for our findings, we compare the results of this study with those from previous investigations, as summarized in Fig.~\ref{fig:logfR0_bar}. \citet{2015PhRvD..92d4009C} employed cluster number counts from the ROSAT Brightest Cluster Sample in combination with external datasets, including CMB, SN, and BAO, obtaining an upper bound of \(-4.79\). \citet{2024A&A...686A.193K} utilized the cross-correlation of galaxies from the BOSS-DR12 survey with CMB lensing data, yielding an upper bound of \(-4.61\). The most stringent constraint to date is reported by \citet{2024arXiv240913556V}, who analyzed SPT galaxy clusters along with CMB data to derive an upper limit of \(-5.32\). Additionally, a recent study by \citet{2024arXiv240915640A} using the full-shape galaxy power spectrum from the BOSS-DR12 dataset provided an upper limit of \(1.53 \times 10^{-5}\) on \(|f_{R_0}|\) in linear space.

In this work, we demonstrate that cosmic shear data can significantly contribute to constraining \(f(R)\) gravity when combined with CMB and BAO datasets. However, the effectiveness of cosmic shear data is currently limited by challenges in data vector modeling and signal noise. Even with the inclusion of blue shear data and the omission of scale cuts, the constraints within the \(f(R)\) framework do not show substantial improvement. While our results are not the most stringent, they are comparable to constraints from previous studies, emphasizing the potential of cosmic shear as a valuable tool for probing \(f(R)\) gravity. This study underscores the importance of incorporating cosmic shear data into analyses of modified gravity and lays the foundation for future refinements in this field.

\section{Conclusion}
In this study, we investigate the potential of cosmic shear data to constrain the Hu-Sawicki \(f(R)\) gravity model, using three Stage-III weak lensing surveys: DES-Y3, KiDS-1000, and HSC-Y3. By employing the nonlinear matter clustering predictions provided by the FREmu emulator, which was trained on Quijote-MG simulations, we perform a comprehensive analysis of cosmic shear constraints and explore the parameter space of \(f(R)\) gravity.

Our results indicate that cosmic shear data alone place limited constraints on the modified gravity parameter \(\log_{10}|f_{R_0}|\). However, when combined with external datasets, including the PR4 CMB and DESI BAO, cosmic shear measurements significantly enhance the overall constraints, yielding a robust upper limit of \(\log_{10}|f_{R_0}| < -4.98\) at the 95\% confidence level. This improvement highlights the complementarity between weak lensing and external probes, where the sensitivity of cosmic shear to nonlinear growth amplifies the precision provided by large-scale measurements in external datasets.

We also emphasize the current limitations and challenges associated with cosmic shear datasets in testing modified gravity, particularly the inability to constrain \(\log_{10}|f_{R_0}|\) in shear-only analyses. These findings underscore the necessity of forthcoming Stage-IV surveys, such as LSST \citep{2019ApJ...873..111I} and Euclid \citep{2022A&A...662A.112E}, which are expected to deliver significantly improved precision and expanded redshift coverage to address these challenges.

Our future work will focus on enhancing the emulator by incorporating galaxy clustering information into the analysis. This will involve extending its predictive capabilities to encompass galaxy clustering signals, such as $3 \times 2$pt statistics, and the full-shape galaxy power spectrum within the framework of $f(R)$ gravity.

In conclusion, this work underscores the critical importance of joint analyses in testing $f(R)$ gravity and provides a framework for future investigations. The synergy of next-generation surveys and advanced modeling techniques will be essential in advancing our understanding of gravity and its role in the evolution of cosmic structures.

\section*{acknowledgments}

We thank Prof. Baojiu Li for inspiring discussions and helpful comments. J.-Q.X. is supported by the National Natural Science Foundation of China, under grant Nos. 12473004 and 12021003, the National Key R\&D Program of China, No. 2020YFC2201603, the China Manned Space Program with grant Nos. CMS-CSST-2025-A01 and CMS-CSST-2025-A04, and the Fundamental Research Funds for the Central Universities. G.-B.Z. is supported by National Key R\&D Program of China No. 2023YFA1607803, NSFC grant 11925303, and by the CAS Project for Young Scientists in Basic Research (No. YSBR-092), the China Manned Space Project with No. CMS-CSST-2021-B01, and the New Cornerstone Science Foundation through the XPLORER prize.


%





\bibliography{sample631}{}
\bibliographystyle{aasjournal}



\end{document}